\documentclass[conference]{IEEEtran}
\IEEEoverridecommandlockouts
\usepackage{cite}
\usepackage{amsmath,amssymb,amsfonts}
\usepackage{algorithmic}
\usepackage{graphicx}
\usepackage{textcomp}
\usepackage{xcolor}
\def\BibTeX{{\rm B\kern-.05em{\sc i\kern-.025em b}\kern-.08em
    T\kern-.1667em\lower.7ex\hbox{E}\kern-.125emX}}

\usepackage[normalem]{ulem}
\newcommand\bfit[1]{\textbf{\textit{#1}}}
\newcommand\bful[1]{\textbf{\uline{#1}}}
\usepackage{hyperref}

\newcommand{\cc}{\textit{Two-Chains}}
\newcommand{\rxid}{\textit{Local Function}}
\newcommand{\txl}{\textit{Injected Function}}
\newcommand{\ssum}{\textit{Server-Side Sum}}
\newcommand{\indput}{\textit{Indirect Put}}

\usepackage{tikz}
\newcommand\copyrighttext{
  \footnotesize \textcopyright 2021 IEEE. Personal use of this material is permitted. Permission from IEEE must be obtained for all other uses, in any current or future media, including reprinting/republishing this material for advertising or promotional purposes, creating new collective works, for resale or redistribution to servers or lists, or reuse of any copyrighted component of this work in other works.}

\newcommand\copyrightnotice{
\begin{tikzpicture}[remember picture,overlay]
\node[anchor=south,yshift=10pt] at (current page.south) {\fbox{\parbox{\dimexpr\textwidth-\fboxsep-\fboxrule\relax}{\copyrighttext}}};
\end{tikzpicture}
}

\begin{document}

\title{Two-Chains: High Performance Framework for Function Injection and Execution}

\makeatletter
\newcommand{\linebreakand}{
  \end{@IEEEauthorhalign}
  \hfill\mbox{}\par
  \mbox{}\hfill\begin{@IEEEauthorhalign}
}
\makeatother

\author{\IEEEauthorblockN{Megan Grodowitz}
\IEEEauthorblockA{\textit{Arm Research} \\
Austin, TX \\
megan.grodowitz@arm.com}
\and
\IEEEauthorblockN{Luis E. Peña}
\IEEEauthorblockA{\textit{Arm Research} \\
Austin, TX \\
luis.epena@arm.com}
\and
\IEEEauthorblockN{Curtis Dunham}
\IEEEauthorblockA{\textit{Arm Research} \\
Austin, TX \\
curtis.dunham@arm.com}
\linebreakand
\IEEEauthorblockN{Dong Zhong}
\IEEEauthorblockA{\textit{The University of Tennessee} \\
Knoxville, TN \\
dzhong@vols.utk.edu}
\and
\IEEEauthorblockN{Pavel Shamis}
\IEEEauthorblockA{\textit{Arm Research} \\
Austin, TX \\
pavel.shamis@arm.com}
\and
\IEEEauthorblockN{Steve Poole}
\IEEEauthorblockA{\textit{Los Alamos National Laboratory} \\
Los Alamos, NM \\
swpoole@lanl.gov}
}

\maketitle
\copyrightnotice

\begin{abstract}
Some important problems, such as semantic graph analysis, require large-scale
irregular applications composed of many coordinating tasks that operate on a
shared data set so big it has to be stored on many physical devices. In these
cases, it may be more efficient to dynamically choose where code runs as the
applications progresses. Many programming environments provide task migration
or remote function calls, but they have sharp trade-offs between flexible
composition, portability, performance, and code complexity.

We developed \cc{}, a high performance framework inspired by active message
communication semantics. We use the GNU Binutils, the ELF binary format, and the
RDMA network protocol to provide ultra-low granularity distributed function
composition at runtime in user space at HPC performance levels using C libraries.
Our framework allows the direct injection of function binaries and data to a
remote machine cache using the RDMA network. It interoperates seamlessly with
existing C libraries using standard dynamic linking and load symbol resolution.
We analyze function delivery and execution on cache stashing-enabled hardware
and show that stashing decreases latency, increases message rates, and improves
noise tolerance. This demonstrates one way this method is suited to increasingly
network-oriented hardware architectures.
\end{abstract}

\section{Introduction}\label{sec:intro}
The basis of all current mass-produced computing technology is the RAM-stored
program model. To execute instructions, processors fetch code as stored data
from addressable memory. Executing instructions induces the system to pull in
data and use other resources. In distributed systems, many processors are spread
out among vast memory and resources. So, it is sometimes more efficient to
reverse the pull modality and instead push the instructions where there are
better resources or better data locality. When a system is heterogeneous, some
processors can do some operations more efficiently than others. Therefore, it is
sometimes more efficient to put instructions on a processor with different
capabilities.

Execution placement to increase efficiency has many potential solutions up and
down the hardware/software stack. Active messages~\cite{bonachea2018gasnet},
message-driven distributed objects~\cite{acun2014parallel}, remote procedure
calls~\cite{braam2019lustre}, processing-in-memory~\cite{ahn2015scalable},
lambda functions~\cite{fouladi2019laptop}, and more will provide various ways
for a programmer or runtime to physically move computation around hardware
localities of a distributed system.

In this work, we use remote dynamic linking in combination with RDMA and cache
stashing to flexibly compose and inject function binaries and data, around a
distributed system. We use the standard capabilities of the Linux kernel, the
GNU Binutils, static ELF binary modification, and a high-performance user
space runtime library to handle remote linking and loading.

Our method uses compile time code modifications to compose portable external
name references, then uses standard (POSIX.1-2001) symbol extraction to bind
those references at runtime. The resulting remote linking capability enables an
active message~\cite{von1992active} programming model consisting of a C language
API to enforce canonical symbolic naming suitable for runtime introspection. A
build toolchain processes C source files, then statically modifies the assembly
to insert a few hooks needed by the runtime.

The runtime is implemented as a plugin to the UCX communication
framework~\cite{shamis2015ucx} that packages and injects active messages over
RDMA using one-sided remote memory operations that trigger execution upon
arrival. The programming model uses \bfit{Two} types of \bfit{C}ooperatively
\bfit{H}andled \bfit{A}ctively \bfit{I}ntegrated \bfit{N}atively
\bfit{S}hared-objects, so we called it the \cc{} framework and this is how it is
referenced in the rest of the paper.

\subsection{Contributions of the Work}
This work contributes an active message framework design and implementation that
is high-performant and interoperable with other libraries. We designed this as a
library extension with the bottom up goal of being able to do one-sided-put with
an encapsulated C function over an RDMA network and have it trigger execution on
the receiver.

We have also designed and implemented the unique feature of embedding code
binaries (functions) within an active message to take advantage of hardware that
stashes data and instructions directly into processor caches.

To make the framework usable, we developed an active messaging interface and
runtime design for a remote linking active message programming environment. The
\cc{} framework presents a unique division of objects into 1) heavyweight shared
libraries used to setup interfaces and synchronize namespaces between processes
and 2) lightweight encapsulated active messages used to push and run code
on-demand over the network. Due to space constrains we do not include the API
description in the paper. We are in process of open sourcing our code under the
UCX GitHub repository~\cite{openucx-website}.

Unlike other task frameworks, \cc{} does not come with much extra baggage like
schedulers or name registries. It packages, transfers, and executes C functions
between processes really fast and clean. It depends only on the UCX communication
library. This is why it is so interoperable with other languages and runtimes.

To demonstrate the benefits, we implemented several benchmarks using the \cc{}
framework and analyzed the performance in several scenarios: \cc{} framework
performance overhead, performance comparison with and without code binaries
included in messages, comparison of performance improvements with hardware
that provides cache stashing, evaluation of tail-latency performance, and the
evaluation of hardware assistance to improve CPU-cycle efficiency in spin waits.

There are definitely security implications of writing programs that use
disaggregated namespaces, runtime symbol resolution, and code movement. A
complete exploration of securing this implementation is outside the scope of
this work. For now, we provide a number of design options to improve security
and discuss issues specific to the motivating application environments where
\cc{} is designed to work well.

The rest of the paper organized as follows.
Section~\ref{sec:background} discusses the existing state of the art in the area
of dynamic computation movement and how \cc{} provides unique capabilities.
Section~\ref{sec:distlink} presents the software architecture design for remote
linking and messaging. Section~\ref{sec:model} presents the \cc{} programming
model, build toolchain, and a UCX library extension to use \cc{} in distributed
applications. Section~\ref{sec:security} discusses the security
implications, opportunities and potential solutions. Section~\ref{sec:methods}
describes our testbed design and active message benchmark programs.
Section~\ref{sec:results} presents and analyses the performance characteristics
of \cc{}.
\section{Motivations \& Background}\label{sec:background}
In general, compute and data migration may help in cases where one process
provides a better resource than another: faster data access, specialized
acceleration, less contention, lower power, etc. Nevertheless, compute and data
migration introduces overheads in the way of task movement~\cite{landsberganalyzing}
or management~\cite{schwarzrock2019influence}.

In particular, we are motivated by data-intensive irregular applications, where
compute migration is more likely to help if:
\begin{itemize}
\item There are unordered concurrent shared writes to arbitrary locations
  throughout a large data set, so bottlenecks result from data sharing across
  memory zones~\cite{sahneh2019racer} or between servers~\cite{xie2019Pragh}.
\item The computation is highly decomposed into small tasks and displays dynamic
  levels of parallelism, so task movement is cheap and load balance is
  tenuous~\cite{Klinkenberg2020Chameleon}.
\item The results of unordered concurrent writes to shared data during
  computation determine code execution paths and runtime levels of parallelism
  and therefore behavior is determined by data races making the application be
  sensitive to long tail latencies.
\end{itemize}

Serving the needs of large-scale distributed applications through computation
movement is a problem that motivates simultaneous innovation in hardware,
runtimes, algorithms, and programming models, since these have all been
co-designed for decades with the exact opposite assumption of stationary
computation with moving data. Because of this, solutions to this problem tend to
be overloaded in terms of innovating contributions, making a steep adoption curve.

\subsection{Runtimes \& Programming Models}
Active messages~\cite{von1992active} combine a data payload with executable code
on a receiver, which is what we do here. The novel feature of \cc{} is the
ability to move the code in messages, and then execute it on arrival without any
virtual environment. The GasNET~\cite{bonachea2018gasnet} system provides an API
for registering and invoking active messages. Unlike GasNET and other similar
systems, \cc{} uses name binding instead of function registration and
incorporates the execution code into the message itself. Another downside of
GasNET compared to \cc{} is that GasNET implements a full PGAS model, which comes
with an added overhead of the memory footprint for a global shared heap.

The Snap Microkernel~\cite{snap2019} project provides a platform for remote
procedure calls in the context of network functionality distribution. Like many
of the other computation placement and migration frameworks, it is a heavyweight
multifunction entity. Our solution could be used as a building block as part of
such a system. In the datacenter setting, lightweight container launch for
Lambda functions is implemented with Firecracker~\cite{agache2020firecracker}.
Another work from Fouladi et al. provides very fast container launch to create
highly granular lambda function execution~\cite{fouladi2019laptop}. None of
these projects addresses issues like heterogeneity of hardware, since
containerization is meant to abstract this. \cc{} can be used as a shim between
hardware and higher level libraries.

Charm++~\cite{acun2014parallel} implements distributed shared objects with the
ability to remotely call methods on those objects. The concepts are similar to
\cc{}, but Charm++ is built on top of other, leaner libraries similar to
\cc{} and UCX, which operate at lower levels of the stack.

The CHAMELEON~\cite{Klinkenberg2020Chameleon} framework by Klinkenberg et al.
uses \texttt{\#pragma}s and runtime APIs to encapsulate OpenMP tasks as
migratable entities in a reactive workload balancer for irregular applications
written in MPI. Unlike that work, \cc{} does not depend on OpenMP or MPI,
does not require using C++, nor requires explicit task progress if the UCX
library uses progress threads. Further, its remote VA resolution process to move
tasks between address spaces is a heavyweight exchange of references via MPI
Send/Recv for each migration event. Our work could potentially be used as a
lower runtime layer to greatly simplify and speed up CHAMELEON, especially since
they found in the course of their work that push-oriented compute movement (as
we have implemented here) is a better mechanism than work stealing for load
balancing since it allows computation-communication overlap.

The FaRM~\cite{dragojevic2014farm} project implements a shared address space
programming model that uses the RDMA network for remote object manipulation. Our
work also uses RDMA, but, in addition, provides the flexibility of moving
user-defined functions and data to remote machines.

\subsection{Network Driven Computing}
Our active message framework shares features with the common practice of
computational offload (e.g.\ code migration) from mobile to cloud/edge
devices~\cite{yousafzai2020process,neto2018uloof,gordon2012comet}
to speed up mobile devices. Unlike the active message work presented here,
mobile code offload does \textit{not} use the code movement itself as
communication. Code offload typically extracts tasks and moves compute to a
heavily virtualized mirror of the mobile device that is just much faster.

This work also resembles user space code patching~\cite{rommel2020global} which
modifies the in-memory process image to replace individual
functions~\cite{neamtiu2006dynamic} or the whole code of a running
application~\cite{hayden2014kitsune} to do software update with no downtime. The
key difference here is that code patching is not fast nor expressive enough to
be a programmable communication and data access model.

\cc{} relies on remote linking for symbol resolution across multiple
processes, like the global namespaces implemented in distributed computing
environments such as Plan 9~\cite{pike1995plan},
Spring~\cite{nelson1994uniform}, and Graphene~\cite{tsai2014cooperation}. But
this toolchain works without needing to bind names through some namespace
manager; it uses ELF library loading as a per-process name resolution mechanism.

\subsection{Hardware for Moving Compute}
Hardware thread co-scheduling to reduce cache line contention is an old
concept~\cite{ousterhout1982scheduling,lozi2012remote} and largely
orthogonal to this work. Recent work of note by Wang et al. modified the NOVA
NVM filesystem to allocate memory mapped file data to specific NUMA nodes and
migrate threads based on memory-as-file accesses~\cite{wang2020numa}. Our work
could interact nicely with such a kernel space modification by putting active
messages payloads into NVM memory-as-files to be picked up by their scheduler.

On the grand scale, the EMU architecture~\cite{dysart2016highly} provides
hardware support to transparently move threads between processors in response to
remote access. \cc{} does not do any automatic function relocation, because
that introduces a very complex scheduling and data layout problem into the
programmer's design space~\cite{rolinger2019optimizing}. However, since \cc{}
can inject functions from process to process with minimal overhead, it
would be an ideal model in which to construct complex, low overhead function
movement policies in userland software.

\section{Remote Linking \& Message Mechanics}\label{sec:distlink}
Dynamic linking and loading has been in standard use since systems became
complex enough to have separately maintained software libraries~\cite{boyer360}.
It allows system updates to many programs at once (without re-linking the
program) by replacing a library at some fixed location or name. Remote runtime
linking extends that concept to allow distributed application updates to
sub-processes of the application that alter subsequent active message behavior
(without re-starting the process) by loading a library into a process to change
the resolution of objects or functions with fixed symbolic names. This way,
applications can implement dynamic control and compute with library loading and
active message linking.

One-sided operations over RDMA networks enable fast, delivery on-demand
execution of active messages. The runtime sets up a receiver thread waiting to
call a function with minimal latency when a message payload arrives, possibly
carrying the function code in the message. For code to travel in the message,
the active message function code has been statically modified to allow runtime
linking against symbols on an arbitrary host by redirecting all \textit{global
offset table} (GOT) accesses to an indirection stored in the message.

\subsection{Reactive Mailboxes}
On-demand active message execution requires 1) a process context for the
incoming message, 2) a destination for messages to arrive, and 3) a mechanism to
execute messages as they arrive asynchronously. Using RDMA operations, we
implement a one-sided mailbox trigger mechanism as shown in
\figurename~\ref{fig:mailbox-diagram}. Memory is pinned for one-sided remote
access by an InfiniBand host adapter and a thread is spawned to wait then wake
up and act when signal values are written to specific mailbox locations. The
thread provides an execution context; pinned memory provides a destination for
data and instruction; triggered wake-up arrival provides an on-demand execution
mechanism.

\begin{figure}[tbp] %
  \centerline{\includegraphics[width=0.4\textwidth]{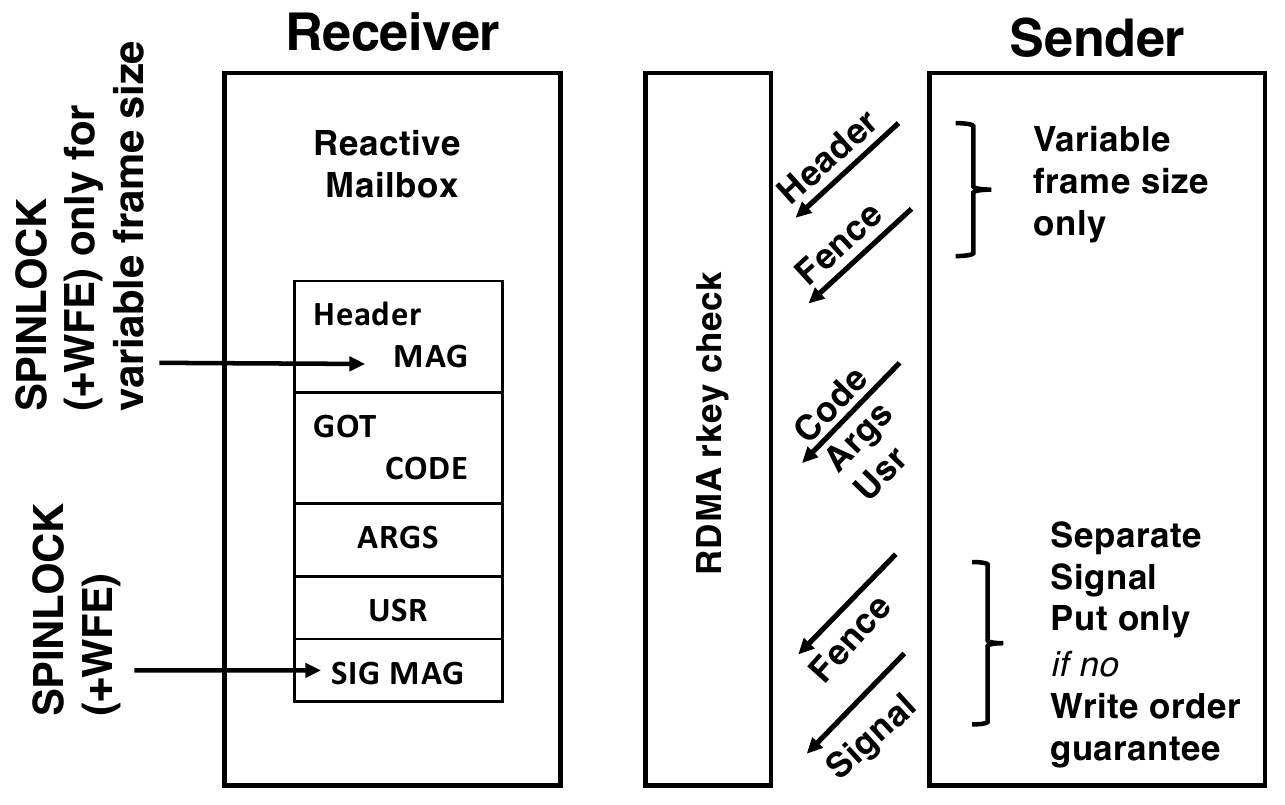}}
  \caption{Message mailbox for on-demand active message}
  \label{fig:mailbox-diagram}
\end{figure}

We do not use interrupts for wake-ups because that would increase latency with
Linux kernel scheduler activity and extra bus transactions. We are targeting
latency sensitive operations with the active message model, so this is not
acceptable. Polling is faster, which is why high-performance communication
libraries like MPI, DPDK, and OVS are moving away from interrupts to use 
polling. The downside of polling is that it uses more energy than interrupts. In
order to mitigate polling inefficiency while preserving its performance we use
hardware-supported sleep operations such as the Arm \textit{Wait For Event} (WFE)
instruction that is set to trigger a wakeup when a mailbox signal location is
altered.

The mailbox expects a specific message format, shown in the left side of
\figurename~\ref{fig:mailbox-diagram}. The \cc{} library provides message packing
routines that call user-provided functions in the active message definition to
setup the data payload. In the current implementation, the message includes a
header, preamble, GOT indirection, code, arguments, and data payload. For the
most compact representation as shown here, we place all code and data together
(USR) and mark all mailbox pages with read, write, and execute permissions. The
runtime can be reconfigured in various ways to separate code and data into
separate locations, make all data read-only, or send no code in messages at all.

RDMA operations for on-demand execution are shown in
\figurename~\ref{fig:mailbox-diagram}, with several different system-dependent
options. If messages use variable-sized frames, the library must be configured
to wait on the final byte of the header (MAG), then retrieve the size, then wait
on the final byte of the message (SIG MAG). If the ordering constraints on RDMA
put operations are not enforced between hosts, then each signal put has to
follow a fence operation and be sent in a separate put from the preceding data
put. Modern servers like the one we use as a testbed for this study enforce
ordering. We also use fixed-size frames for this study, so we can send the
entire message in one put operation.

\subsection{Remote Linking Mechanism}
For code to arrive in messages from the sender process space, we need to resolve
virtual addresses (VAs) for external symbols on the receiver. Linking on message
arrival is far too slow and would require sending a whole library, when we just
want to send a bit of code. At compile time, the binary is modified so that all
references to the \textit{global offset table} (GOT) will redirect through a
pointer stored a fixed PC-relative location that we choose.

This is all supported with standard position-independent library compilation
using \texttt{-fPIC} or \texttt{-fpic} flags and the \texttt{-shared} flag. The
code will be produced with PC-relative addressing modes for accessing data
within the library. Any symbols outside the library are unresolved at compile
time and use indirection through addresses resolved during library load. The
compiler expects the indirection table itself to be loaded a fixed PC-relative
location. To be able to place small code sections at any location on a receiver,
we replace the fixed PC-relative table offsets with indexed access through a
pointer stored at a PC-relative location.

To do this transformation, we force all external symbols to use the GOT by
passing the \texttt{-fnoplt} flag to \texttt{gcc}. Then we find all GOT
indirections in the code and replace them with indirections through a pointer at
a PC-relative location that we choose.

For our test configuration, the GOT redirect is located just before the code in
the message, and is set by the sender after an exchange with the receiver. For
other configurations, the receiver could set a GOT pointer somewhere relative to
the mailbox. The constraint is that the relative location of GOT indirection to
start of code in mailbox must be known at active message compile time.
\section{The \cc{} Toolchain}\label{sec:model}
Over the past year, we have built up the \cc{} toolchain to provide active
messages with functions injected over RDMA. The framework has gone through
several phases, with the current status being close to release as a subcomponent
of the UCX library. \cc{} measures up to the competition with a framework
that:
\begin{itemize}
\item Is not exotic to anyone familiar with C programs.
\item Does not follow an SPMD model for distributed programming,
  e.g.\ MPI. A program can easily define different functions with the same
  symbolic name for different processes, so that when a message arrives it will
  call a function specific to that process, much like function overloading.
\item Encourages single source file active message definitions.
\item Automates source directory-based packaging of active messages.
\item Is interoperable with existing C libraries just as they are, i.e.,
  no re-compilation, no re-linking, no wrappers, etc.
\item Implicitly pulls in read-only data to messages to support functions
  like \texttt{printf}.
\end{itemize}

This environment specifically does not enforce library version dependency any
more than a Linux system enforces what happens when an ELF executable is
dependent on a shared library path and that shared library gets replaced. This
is not a virtual environment; it is subject to the system environment. Since
most applications run under several types of virtualization already --
containers, VMs, hypervisors, etc -- we do not add another layer without strong,
specific reasons.

\subsection{Rieds and Jams}
The \cc{} model defines two types of cooperatively handled actively
integrated natively shared-objects. \bful{R}elocatable \bful{i}nterfac\bful{e}
\bful{d}istributions (rieds) are shared libraries that one process drives over
to some remote process to dynamically setup interfaces and data objects as
needed. Jams are \bful{j}ust \bful{a} \bful{m}obile \bful{s}egment. In the
current work, jams are C functions with payload and fast relocation updates that
a process packs into active message for injections to another process.
In this work, we focus on engineering and performance of jams. We use rieds
only as dynamic libraries that are loaded and auto-initialized in \cc{} packages
to support our benchmarks. Ongoing development will further improve our rieds
for future work.

The \cc{} are organized into packages. Each package has a
package name that is set when the package is built. A package contains elements;
each has a unique element ID and element name
within the package. The current build tools take a list of jams and rieds with
source files located in a subdirectory tree. For now, the build tools expect
each element to be defined in one canonically named source file, e.g.\
\texttt{jam\_append.amc} or \texttt{ried\_array.rdc}.

The build process generates a package header file and shared libraries in the
package install directory. At this time, the build process requires an
installation path to enable simpler runtime APIs for loading package elements.
Once the package is installed, a program includes the generated package header
and the \cc{} runtime headers.

\subsection{Active message invocation methods}
\cc{} supports two types of invocation methods: local function invocation
based on receiver ID (\rxid{}) and function invocation based on binary code in
the message itself (\txl{}). \figurename~\ref{fig:TXL_frame} represents the
\txl{} message layout in memory and \figurename~\ref{fig:RXID_frame} represents
the \rxid{} message layout. Since \txl{} has to update the GOT table and invoke
code communication over the network, the layout in \figurename~\ref{fig:TXL_frame}
includes a patched GOT (GOTP) section and a code section (CODE), in addition to
data payload (USR). Otherwise, the \rxid{} method uses the same message formats,
headers, source code files, header generation, and internal APIs as the \txl{}
mechanism with the difference that it invokes the local function representation
from the library instead of the function code coming over the network.

\begin{figure}[tbp] %
  \centerline{\includegraphics[width=0.4\textwidth]{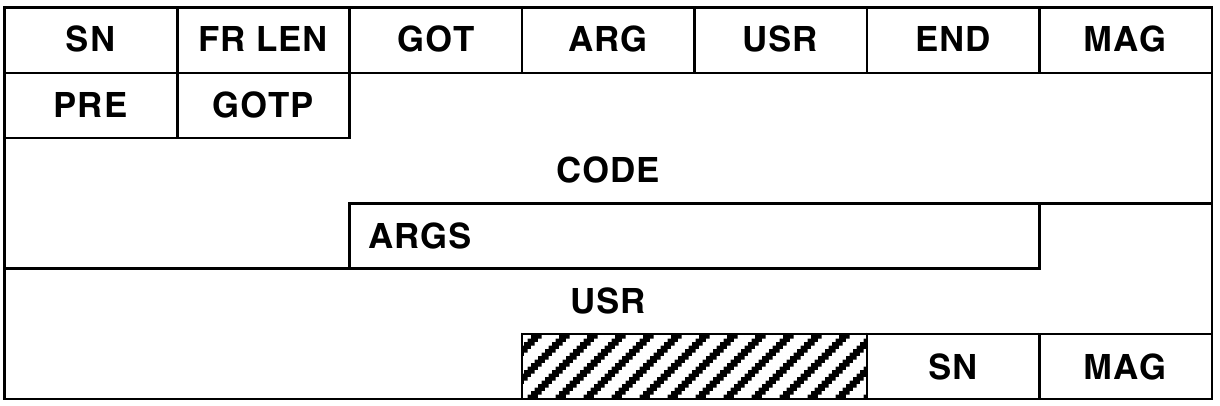}}
  \caption{\txl{} GOT patched from receiver}
  \label{fig:TXL_frame}
\end{figure}

\begin{figure}[tbp] %
  \centerline{\includegraphics[width=0.4\textwidth]{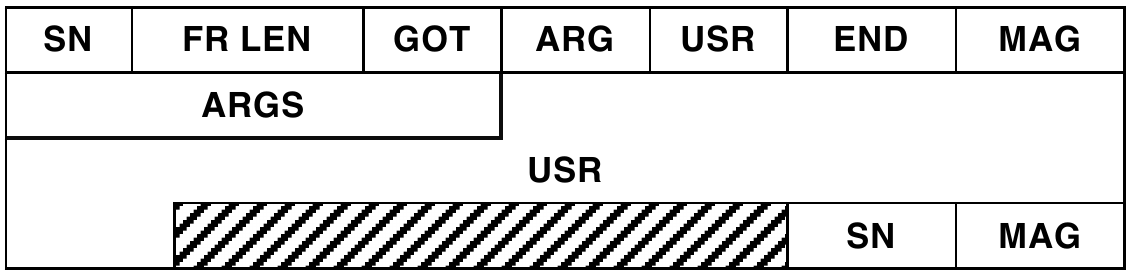}}
  \caption{\rxid{} code in loaded library on receiver}
  \label{fig:RXID_frame}
\end{figure}

The \rxid{} procedure was originally designed as a benchmark to provide a
comparison of the overheads of moving code over network versus sending payload
data and a function ID over network. After implementing it, it was surprisingly
straightforward to provide this option as a core library function.

The toolchain generates a shared library of the same set of active messages
compiled without any modification for GOT patching. The \rxid{} shared library
is loaded on the receiver and provides a vector of function pointers that are
called by using the ID included in the active message header.

By providing both in the same package from the same source, the same code could
be ported between systems where different types provide better performance. It
may also be beneficial to mix message types so that some active messages contain
code and others do not. We are continuing to investigate these types of
tradeoffs between remote linked and locally invoked active message types.
\section{Security Implications and Mitigations}\label{sec:security}
A full security model design and implementation is well beyond the scope of this
paper. This section provides an overview of security challenges and directions
for security improvements.

For our \cc{} framework implementation, we have relied on the built-in
security mechanisms defined by the UCX framework and IBTA
standard~\cite{ibta-specification}, which underpins RDMA interconnects.
Specifically, we are using a remote access key (RKEY) to register and control
remote memory accesses. For IBTA interconnects, the RKEY is defined as a
32-bit value. When the memory is registered for remote memory access, the
underlying interconnect generates the RKEY based on a virtual memory address and
the permissions (remote read, write, or atomic access). In order to access the
memory region over the RDMA interconnect, the target process has to provide the
RKEY to the RDMA initiator through an out-of-band channel. Then, the remote
memory access initiator uses the RKEY to remotely read and write to the target
process memory. If the process accesses the memory with an invalid RKEY, the
request gets rejected at the hardware level.

There are a number of security concerns~\cite{taranov-redmark} regarding the
strength of RKEY protection as defined by the IBTA standard. Improvements to the
IBTA security model are out of scope for this work. However, since we have
constructed this as a module of the UCX framework, the implementation is not as
strictly tied to the IBTA network implementation.

For our primary target use case of massive analytics computations on a large
dataset, the attack surface can be greatly reduced by encasing the entire
application. For sensitive data analysis in government, finance, oil and gas,
etc, this is already done with physically isolated on-site clusters, air-gapping,
etc. Recent work by Zhu et al.~\cite{zhu2020enabling} describes a way to create
an enclave for an analytics computation at the rack level, which has a similar
effect.

Aside from isolating the entire application and data, we have already mentioned
several reconfigurations for active messages that improve security. The
following measures can be employed with straightforward modifications to the
runtime.
\begin{itemize}
	\item Separate the user data payload area from the rest of the message and
	make the function arguments read-only so that the code can be placed on a
	different memory page than writable data, and writable data will not
	reside on executable pages.
\item Do not accept GOT pointer indirection in the message from a sender. Have
	the receiver insert the GOT pointer on message arrival from a secure
	read-only location~\cite{jeong2020cfi}.
\item Keep ried interfaces and/or jam libraries in secure enclaves.
\item Extend the IBTA standard to support executable permissions in addition to
	read, write and atomic.
\end{itemize}

The performance impact of these options is a subject for future study, but none
of these would necessarily incur large performance penalties.

The feasibility of maintaining libraries in enclaves is discussed by Wang et
al.~\cite{wang2020building} in a presentation of their real world experience
porting Rust libraries into enclaves. In terms of large corporate settings, we
would generally expect \cc{} to be used by a platforms team, and not the
product teams. In such a setting, a platform group would work with product
groups to curate a secure, minimal set of rieds and jams to support a variety of
products.
\section{Experimental Design}\label{sec:methods}
\cc{} was designed from the ground up to make active message delivery and
execution as close to an RDMA put operation as possible in performance and
programming effort, even when instructions are delivered in the message. This
fills a niche for general purpose, high performance one-sided active messaging
with few external dependencies. To verify that \cc{} meets the performance goals
we set, we designed and implemented set of jam benchmarks and integrated them
into the existing performance testing tool in the UCX library.

\subsection{Benchmark Shapes}
The performance tool for UCX connects two hosts over a network, and runs a
number of iterations of a particular operation to determine bandwidth, message
rates, and latency of the operation.

\subsubsection{Ping Pong}
The ping pong benchmark shape sends one message at a time between hosts. Each
host has one message mailbox. It waits on the message signal (\textit{ping}),
executes the active message on arrival, then sends a response (\textit{pong})
message to the initiator of the message, which also executes the active message
on \textit{pong} arrival. This shape of the benchmark is used for measuring the
half round-trip (one-way) active message latency.

\subsubsection{Injection Rate}
The injection rate benchmark is used to test the rate at which active messages
can be processed when a sender puts messages on the receiver as fast as possible.
The UCX communication library has mechanisms and buffers for flow control, but,
since the \cc{} runtime already waits on mailbox data to arrive, we used our own
flow control to avoid adding extra overhead to the reactive mailbox. In
injection rate benchmarks, the receiver has \textit{M} banks, where each bank
has \textit{N} mailboxes. The sender keeps one signal flag per bank, which is
reset after sending \textit{N} messages to the matching bank on the receiver.
Every time the receiver empties a bank, it sets the flag for that bank on the
sender. The sender will not send new messages to a bank until the flag for that
bank is set.

\subsection{Benchmark Functions}
Each of the new benchmark shapes can run any active message in the \cc{} test
package that is generated and installed with the UCX performance tester. The
following jams were used for benchmarking the implementation.

\subsubsection{\ssum{}}
The simplest active message here is the \ssum{}. \ssum{} loops over
all of its payload in order to accumulate a sum. Then, it stores the result at
the next spot in an array in the server.

\subsubsection{\indput{}}
The indirect put benchmark models a common distributed use case where a program
wants to access and modify some data structure where every access goes through a
level of indirection. Graph structures and index tables are a prime example. In
~\cite{snap2019} one-sided indirected put is presented as an extension of RDMA
semantics. In the work by Xie et al.~\cite{xie2019Pragh}, an RDMA caching system
is designed to move hash keys and values around at runtime in response to
changing graph requirements.

Using our indirect put operation, as shown in \figurename~\ref{fig:indirect_put},
a client can put an element (\texttt{array[]}) into the server’s memory. Each
element is indexed by an arbitrary key chosen by the client that the server then
uses to probe the index/offset (hash value) and store the data. Once the unique
key is selected, the client issues an active message to the server with the
indirect put request. The active message takes three major steps: (1) It pushes
the key into the hash table, using the key to probe the hash table. (2) It
chooses a proper offset and stores it in the hash table associated with this
particular key, which means the client has full control over the indirect
distribution of the data and the lookup function itself. (3) Finally, the server
issues a memory copy to store the payload data into its memory by copying the
payload data into the \texttt{put\_count * type size} bytes starting at
\texttt{base address + offset}.

\begin{figure}[tbp] 
  \centerline{\includegraphics[width=0.5\textwidth]{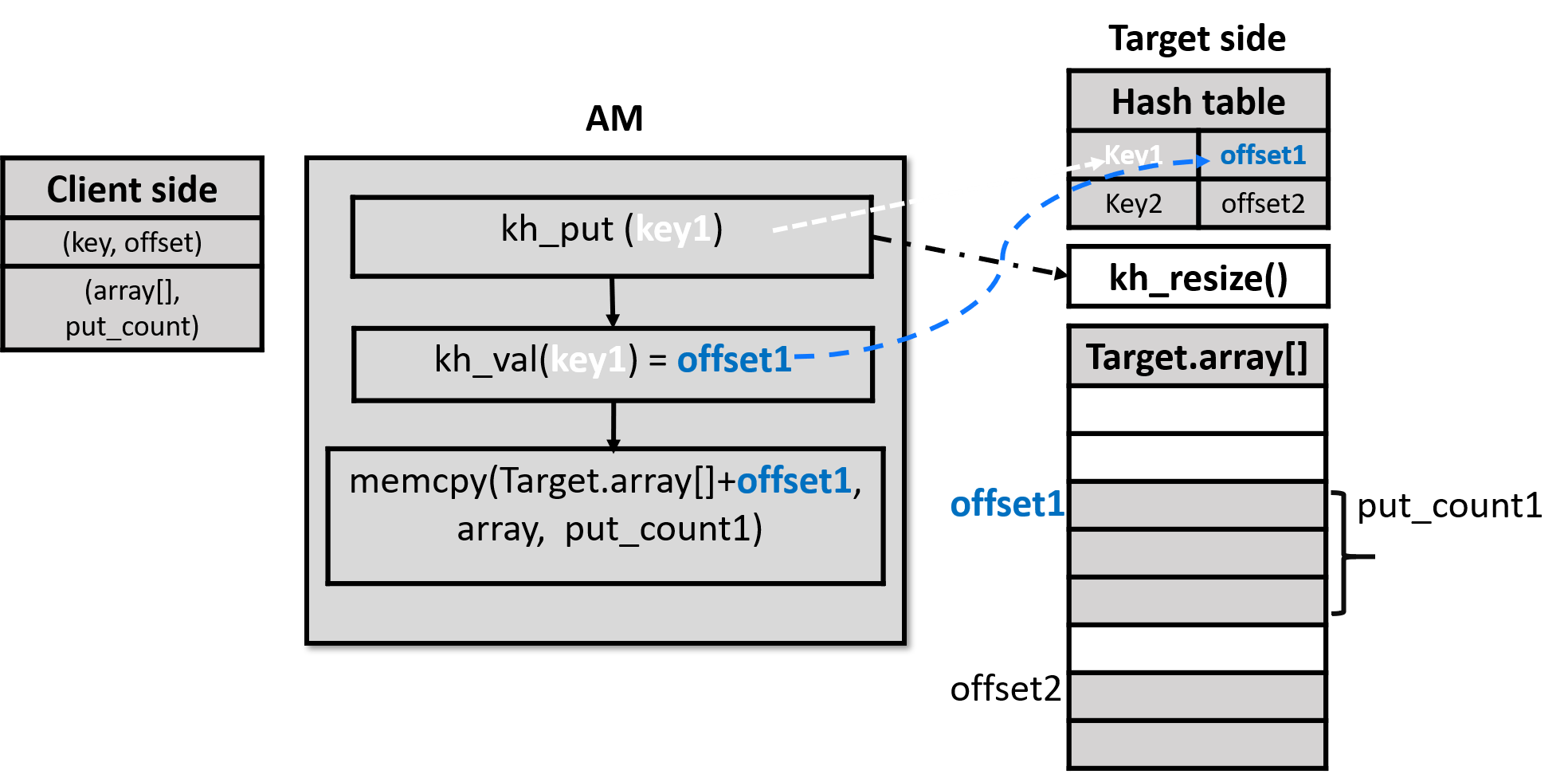}}
  \caption{Indirect Put active message (AM) function.
  The function calculates an offset (\texttt{offset1}) based on the key
  (\texttt{key1}) and copies the data (\texttt{array[]}) to the offset location.}
  \label{fig:indirect_put}
\end{figure}

\subsection{Testbed Platform}
The development and evaluation testbed for this work consisted of two servers,
each with a 4-core, Arm-based modern superscalar processor with a 1MB dedicated
L2 cache per core, a 1MB shared L3 cache per 2-core cluster, and a 8MB shared
last level cache (LLC). The core clock is 2.6GHz and the on-chip interconnect
clock is 1.6GHz. Each server has 16GB of DDR4-2666 main memory. For the
interconnect feeding RDMA writes from the network to the test systems, we used
two Mellanox/Nvidia ConnectX-6 200Gb/s InfiniBand dual-port HCAs. On each machine, the
HCA was plugged into a PCIe Gen4 slot. The two systems were connected back-to-back
(no InfiniBand switch) using the first port on each ConnectX-6 HCA. The second
port of each HCA was not used.

LLC-stashing is supported in these systems. The PCIe root complex controlling
the ConnecX-6 HCA is connected into the on-chip interconnect. Traffic arriving
from the network is stashed into the LLC and, eventually, written back to the 
main memory.

We are interested in cache stashing because it is a technique to minimize
cache misses when data are arriving from the network. Since \cc{} injects data 
and instructions over the network into remote machines, we expect stashing to
provide better network performance characteristics.

To enable hardware-feature comparisons, each server's firmware was configured
to allow toggling of the LLC-stashing and the prefetching mechanisms. The
servers used Fedora 30, running a custom Linux 5.4 kernel, modified to allow
user space control of the CPU prefetching mechanisms. We used the RDMA and
InfiniBand drivers that came with the kernel, versioned \texttt{20.1-3.fc30}.
\section{Performance Results and Analysis}\label{sec:results}
Rather than targeting another active message implementation as a performance
baseline, we compared \cc{} against RDMA puts. We are interested in maximizing
the benefits of low-latency, high-bandwidth one-sided communication from the
reactive mailbox design. There are no other stand-alone active message
frameworks that we know of that deliver and execute active messages in this way.

First, we verified that the \cc{} framework and its reactive mailbox were not
introducing any performance penalties compared to UCX's put operation.
\figurename~\ref{fig:noex-lat} and \figurename~\ref{fig:noex-bw} compare the
latency and bandwidth characteristics of UCX put with those of the active
message put when running in the without-execution configuration. Running in the
without-execution configuration uses \cc{} to inject active messages, deliver
them to the mailboxes and trigger their arrival, but skips the actual function
invocation. From \figurename~\ref{fig:noex-lat} and \figurename~\ref{fig:noex-bw},
we see no significant drop in latency, 1.5\% at worst, for messages going to the
\cc{} reactive mailboxes. In fact, we see bandwidth improvement across all
message sizes tested when running in the \cc{} without-execution configuration,
ranging from a 1.79× speedup up to a 4.48× speedup. Our mailbox flow control and
memory actually see improvement because the standard UCX put operation has more
library overhead for flow control and detecting message completion.

\begin{figure}[tbp] 
  \centerline{\includegraphics[width=0.5\textwidth]{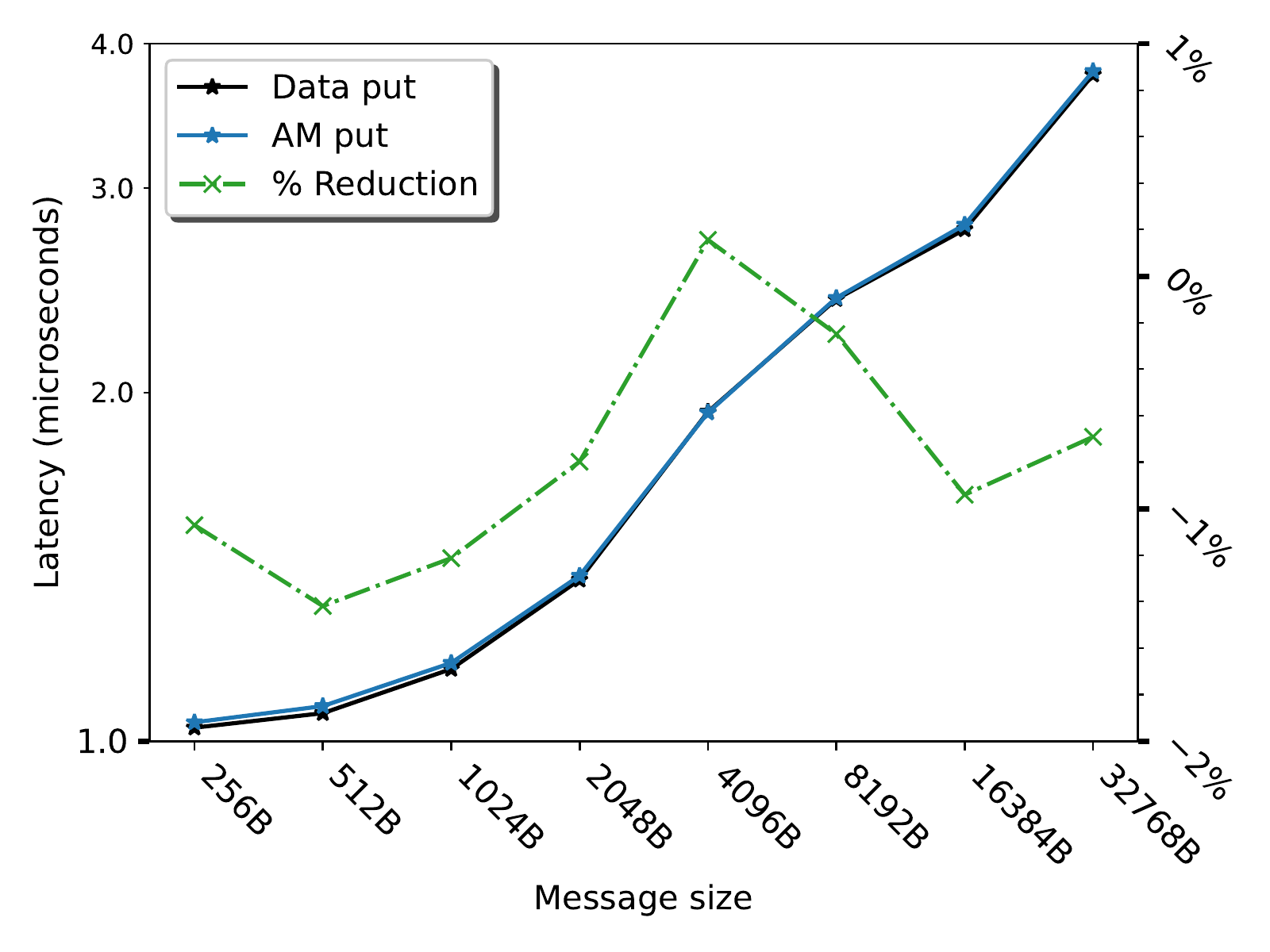}}
  \caption{\ssum{}: \cc{} active message (AM) put without-execution latency overhead}
  \label{fig:noex-lat}
\end{figure}

\begin{figure}[tbp] 
  \centerline{\includegraphics[width=0.5\textwidth]{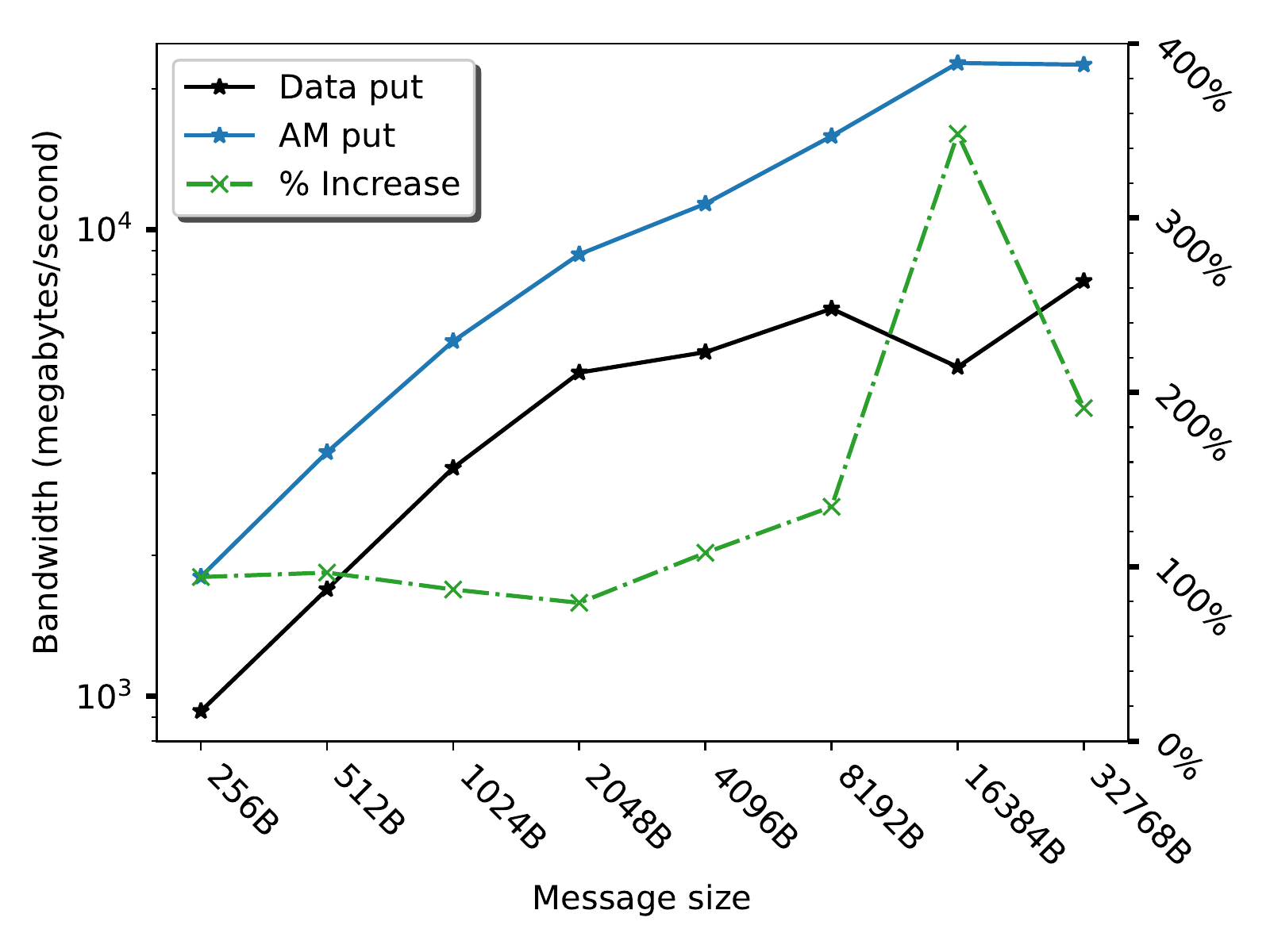}}
  \caption{\ssum{}: \cc{} active message (AM) put without-execution bandwidth overhead}
  \label{fig:noex-bw}
\end{figure}

\subsection{Comparison of Local versus Injected Function Invocations}
The goal of this experiment is to evaluate the trade-offs of \rxid{} versus
\txl{} invocation for various message sizes. Since \cc{} lets code reside on the
sender or receiver, we can measure the overhead of \txl{} execution relative to
\rxid{} execution.
\figurename~\ref{fig:loc-lat} and \figurename~\ref{fig:loc-mr} contrast the
latencies and message rates for \txl{} and \rxid{}. Note that the comparison is
between equivalent payload, but not equal message size. The code for \indput{}
is 1408 bytes when shipped, and messages are sized to the nearest 64B: the
1-integer message size is 64B for \rxid{} versus 1472B for \txl{}. The
approximately 40\% losses in latency and bandwidth correspond to sending much
more data per \txl{} message for small payloads. Once the payload is large
enough, the overhead of moving code becomes negligible, as expected.

\begin{figure}[tbp] 
  \centerline{\includegraphics[width=0.5\textwidth]{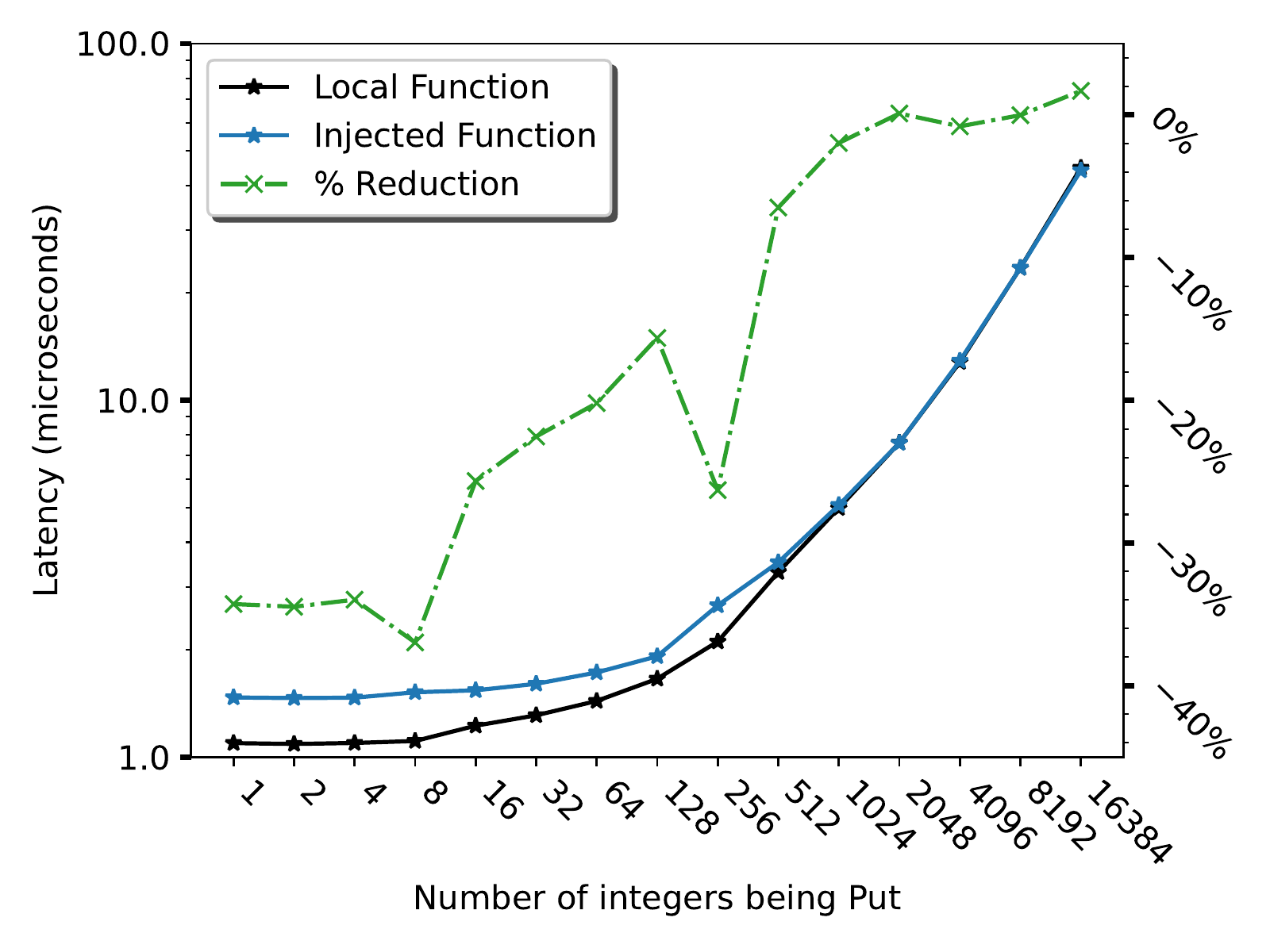}}
  \caption{\indput{}: Latency comparison between injected and local function invocation}
  \label{fig:loc-lat}
\end{figure}

\begin{figure}[tbp] 
  \centerline{\includegraphics[width=0.5\textwidth]{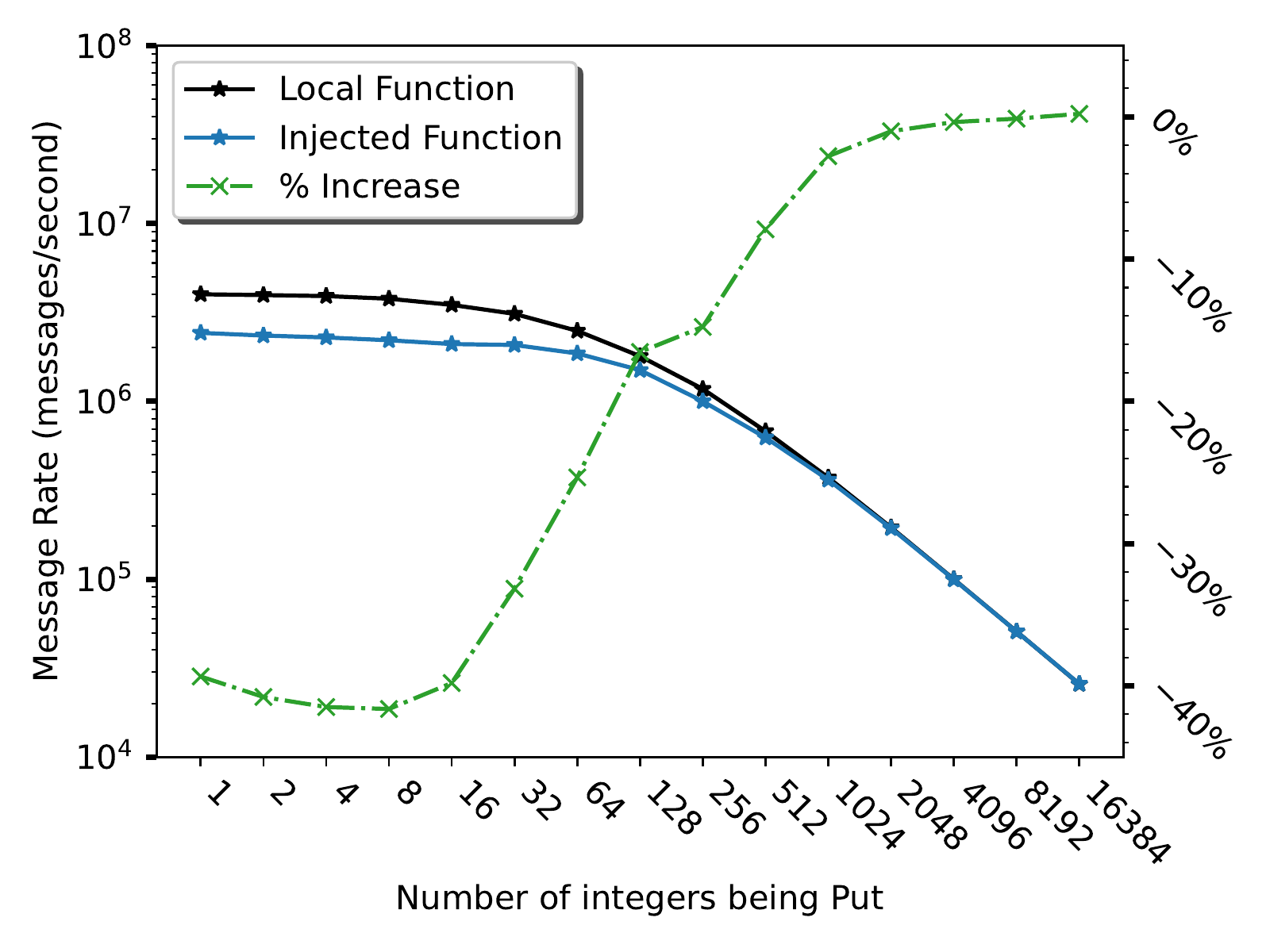}}
  \caption{\indput{}: Message rate comparison between injected and local function invocation}
  \label{fig:loc-mr}
\end{figure}

The irregularities in the latency difference at 8 and 256-integer payload sizes
are an interesting and relevant artifact from using the UCX framework. For
efficient performance, the UCX library will change the protocols used for sending
a message based on its size. When a message is just over the threshold size to
move into a new code path, there will be a slight performance degradation since
the message size is just within the acceptable range. For each of the 8 and
256-integer payloads, the \txl{} message has just crossed a new code path
threshold since it is larger, which is slightly less performant at that message
size. \ssum{} results are similar to \indput{}, except that the code is smaller
so the convergence to 0\% performance loss happens sooner, with \ssum{} around
the 64-integer scenario while \indput{} goes all the way to 1024 integers.

\subsection{Latency \& Message Rate Improvement with Cache Stashing}
As we mentioned earlier, in our implementation of the active message protocol we
take advantage of cache stashing capabilities available in the system. In order
to disable or enable stashing to LLC we use low-level controls that let us
explicitly disable or enable stashing for the ConnectX-6 device.
\figurename~\ref{fig:indput_stash_lat} shows a decrease in latency for
the \indput{} benchmark when cache stashing is enabled. Stashing the message
code and data to LLC improves latency by up to 31\%. However, once the message
size is large enough to trigger the prefetcher to start pulling the message data
on arrival, the difference in latency for messages going to DRAM versus LLC
starts narrowing, as prefetches are issued ahead enough to mask the larger DRAM
access latency. The trends for latency improvements are similar for the
\ssum{} benchmark, which we omit for space.

\begin{figure}[tbp] 
  \centerline{\includegraphics[width=0.5\textwidth]{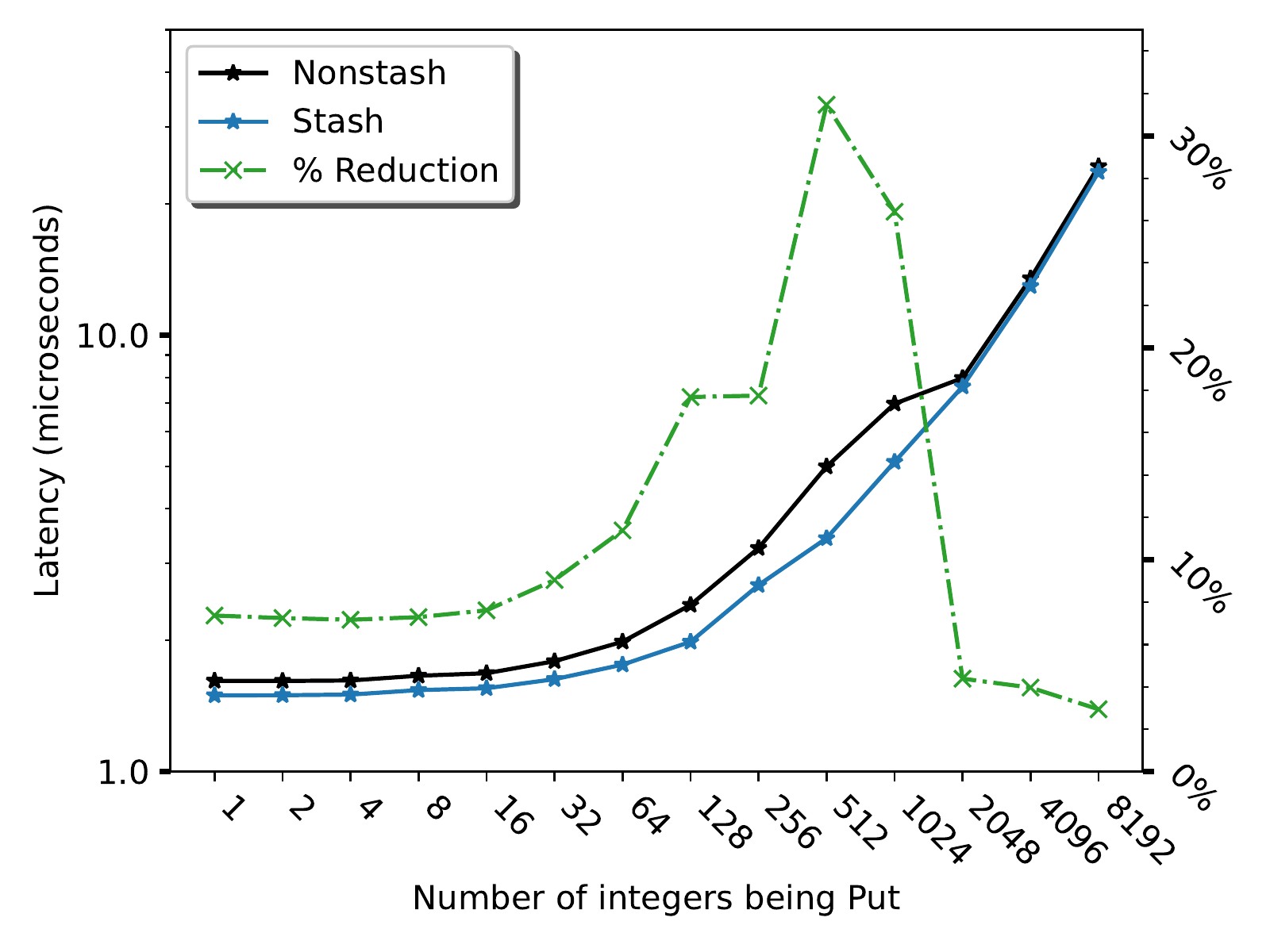}}
  \caption{\indput{}: Latency reduction with LLC stashing enabled (Stash) versus LLC stashing disabled (Nonstash)}
  \label{fig:indput_stash_lat}
\end{figure}

The benefits of LLC stashing for \indput{} message rates
(\figurename~\ref{fig:indput_stash_mr}) are clear. For \indput{}, there is a 92\%
(1.9×) message rate increase for small put counts, with this advantage reducing
as message sizes get large enough to benefit from the prefetcher. The small code
footprint and easy-to-prefetch linear access pattern displayed by \ssum{} result
in up to 28\% improvement for message rates at any size (plot omitted for space).

\begin{figure}[tbp] 
  \centerline{\includegraphics[width=0.5\textwidth]{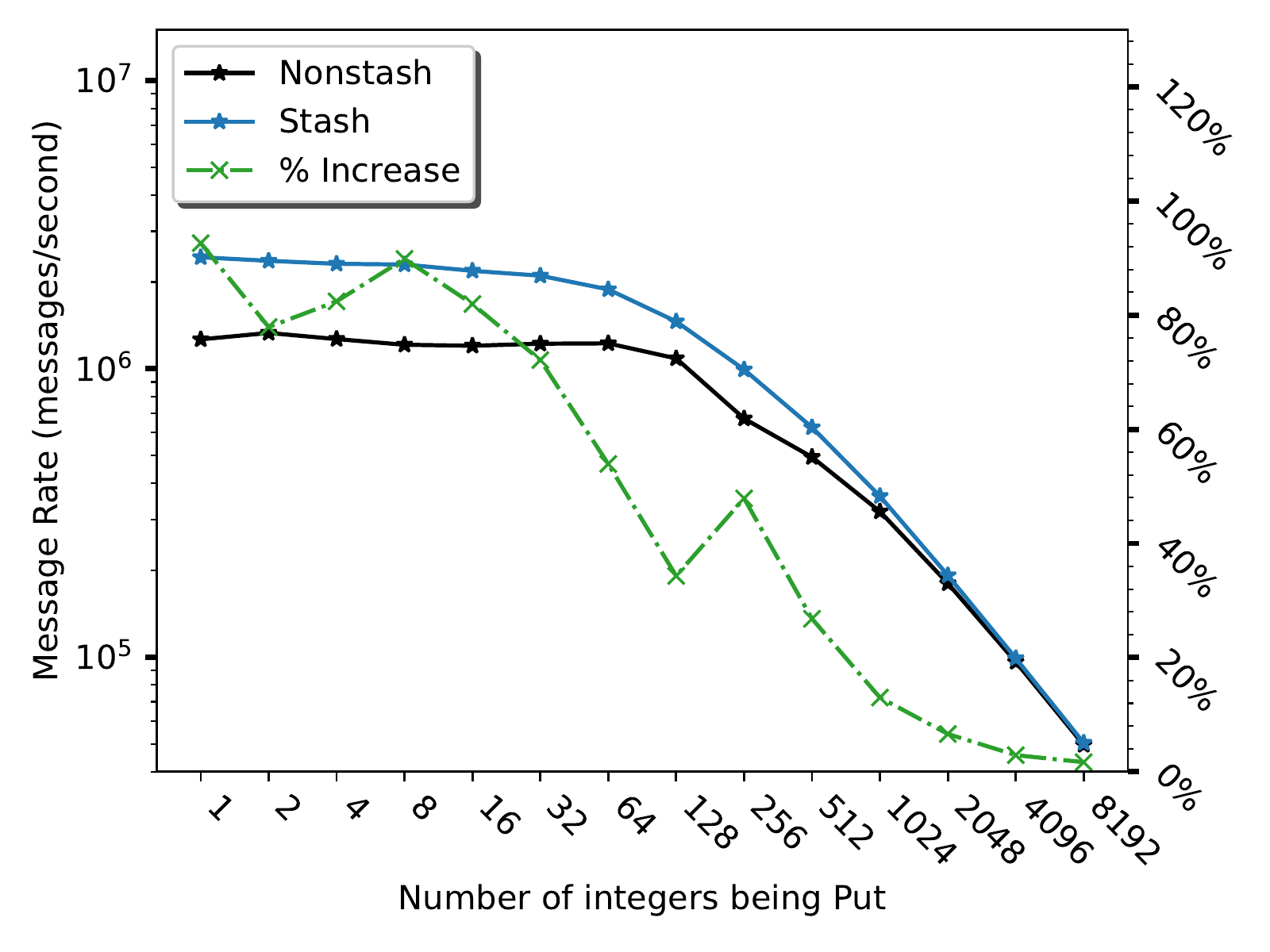}}
  \caption{\indput{}: Message rate increase with LLC stashing enabled (Stash) versus LLC stashing disabled (Nonstash)}
  \label{fig:indput_stash_mr}
\end{figure}

These results are limited in scope due to the test harness that we have
employed. But they do make the larger point that the stashing or RDMA'ing of
functions involves the entire cache hierarchy. Applications that would benefit
the most from being able to stash encapsulated functions would be those that are
to prefetch or those that have large reuse distance within the application and
are likely to be cold in the memory hierarchy when invoked.

\subsection{Tail Latency with Stashing}
Our active message runtime does not take over the entire system; in a realistic
scenario, it would run alongside other applications to minimize resource
underutilization, the system would be run as close to capacity as possible. To
quantify the adverse effects of running at-capacity systems, we measured the
tail latency of \cc{} active messages.

To test the characteristics of LLC stashing \cc{} active messages on overloaded
systems, we ran \texttt{taskset -c 0-3 stress-ng --class vm --all 1} along with
the benchmark to generate stressful work for the paging and memory systems. One
way to detect service degradation is to measure the tail latency of the active
messages, so, along with 50\textsuperscript{th} percentile (typical or median)
latency, we collected the 99.9\textsuperscript{th} percentile (tail) latency. We
also calculated what we call \textit{tail latency spread} which is how much
larger the tail latency is than the median and is show in \eqref{eq:spread}.

\begin{equation}
Latency_{tail} spread = \frac{Latency_{tail} - Latency_{typical}}{Latency_{typical}}
\label{eq:spread}
\end{equation}

The smaller tail latency spread is, the narrower the latency distribution is. A
narrower latency distribution leads to a more predictable and less erratic
latency. The tail latency spread can be thought of as a sort of latency
\textit{variance}. We plotted tail latency, typical latency and tail latency
spread below.

\figurename~\ref{fig:overload_indput} and \figurename~\ref{fig:overload_ssum}
show the latencies and tail latency spread of executing the \indput{} and
\ssum{} benchmarks on an overloaded system. The \indput{} 
(\figurename~\ref{fig:overload_indput}) tail latency is up to 2.4× better when
LLC stashing is enabled. With stashing, the tail latency spread peaks at 182\%,
while non-stashing has an erratic behavior. This erratic behavior is likely due
to interference by the stress workload competing for memory access. By writing
directly to the LLC, stashing reduces active message memory bandwidth
utilization.

\begin{figure}[tbp] 
  \centerline{\includegraphics[width=0.5\textwidth]{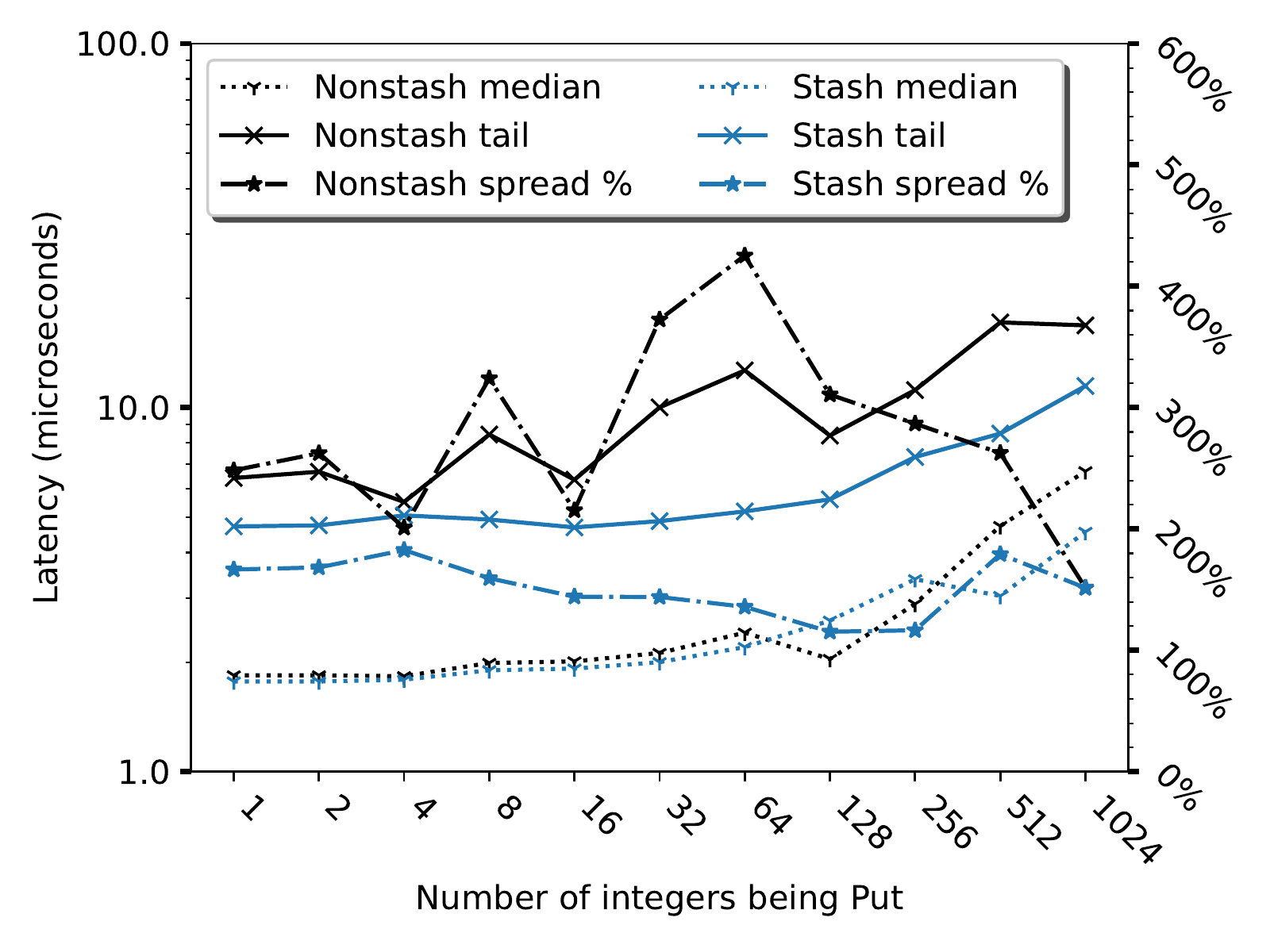}}
  \caption{\indput{}: LLC stashing enabled (Stash) and disabled (Nonstash) latency on fully-loaded system}
  \label{fig:overload_indput}
\end{figure}

As shown in \figurename~\ref{fig:overload_ssum}, the \ssum{} LLC stashing
99.9\textsuperscript{th} tail latency is generally better than that of the
non-stashing scenario, in some cases performing twice as fast. Starting with the
2KB message size, stashing provides a tighter latency distribution compared to
the non-stashing case, with a tail latency no larger than 137\% of the median
latency. These results highlight the benefits of stashing even on at-capacity
systems.

\begin{figure}[tbp] 
  \centerline{\includegraphics[width=0.5\textwidth]{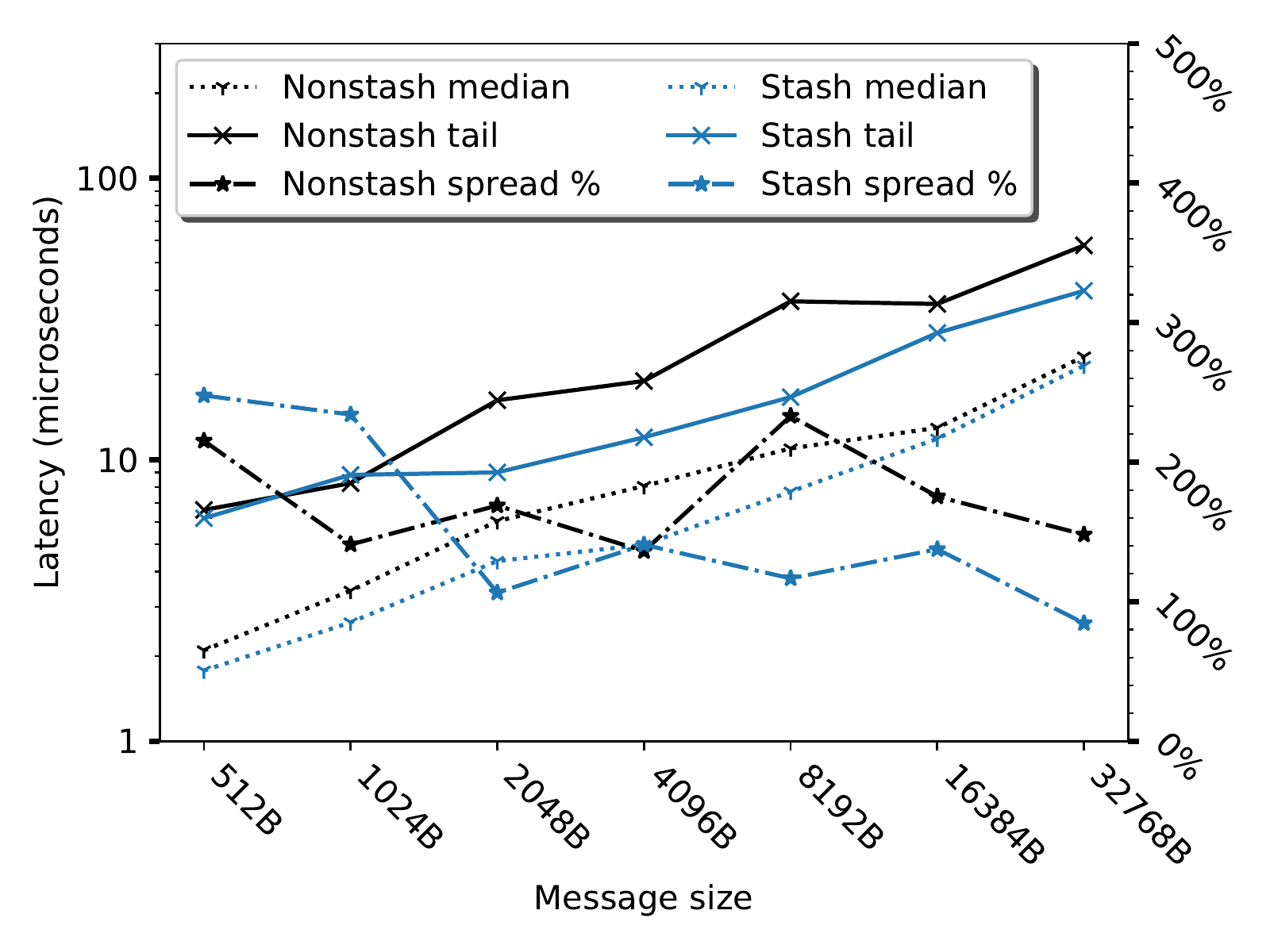}}
  \caption{\ssum{}: LLC stashing enabled (Stash) and disabled (Nonstash) latency on fully-loaded system}
  \label{fig:overload_ssum}
\end{figure}

\subsection{Efficient Spin Polling}
In section \ref{sec:distlink}, we described the \cc{} implementation of
efficient message polling. We use Arm's \textit{WFE} instruction to reduce the
number of CPU cycles the framework spends on spin-poll waiting on an active
message arrival signal. To see the effects on cycles, we gathered the CPU cycles
counters for the full benchmark run, including 10,000 warm up iterations and
1,000,000 latency-measuring iterations.

\figurename~\ref{fig:wfe_ind} shows the active message latency and the
full-runtime CPU cycle counters for our \indput{} benchmark. The latency
remains the same for most payload sizes tested when \textit{WFE} is inserted in
the wait loop. Compared to busy-waiting (\textit{Polling}), see up to 1.5\%
latency penalty, at the 64B data payload size. When \textit{WFE} is used, we see
between a 3.8× and 2.5× CPU cycle reduction. The cycle-count reduction comes
solely from the waiting-for-active message portion of the code.

\begin{figure}[tbp] 
  \centerline{\includegraphics[width=0.5\textwidth]{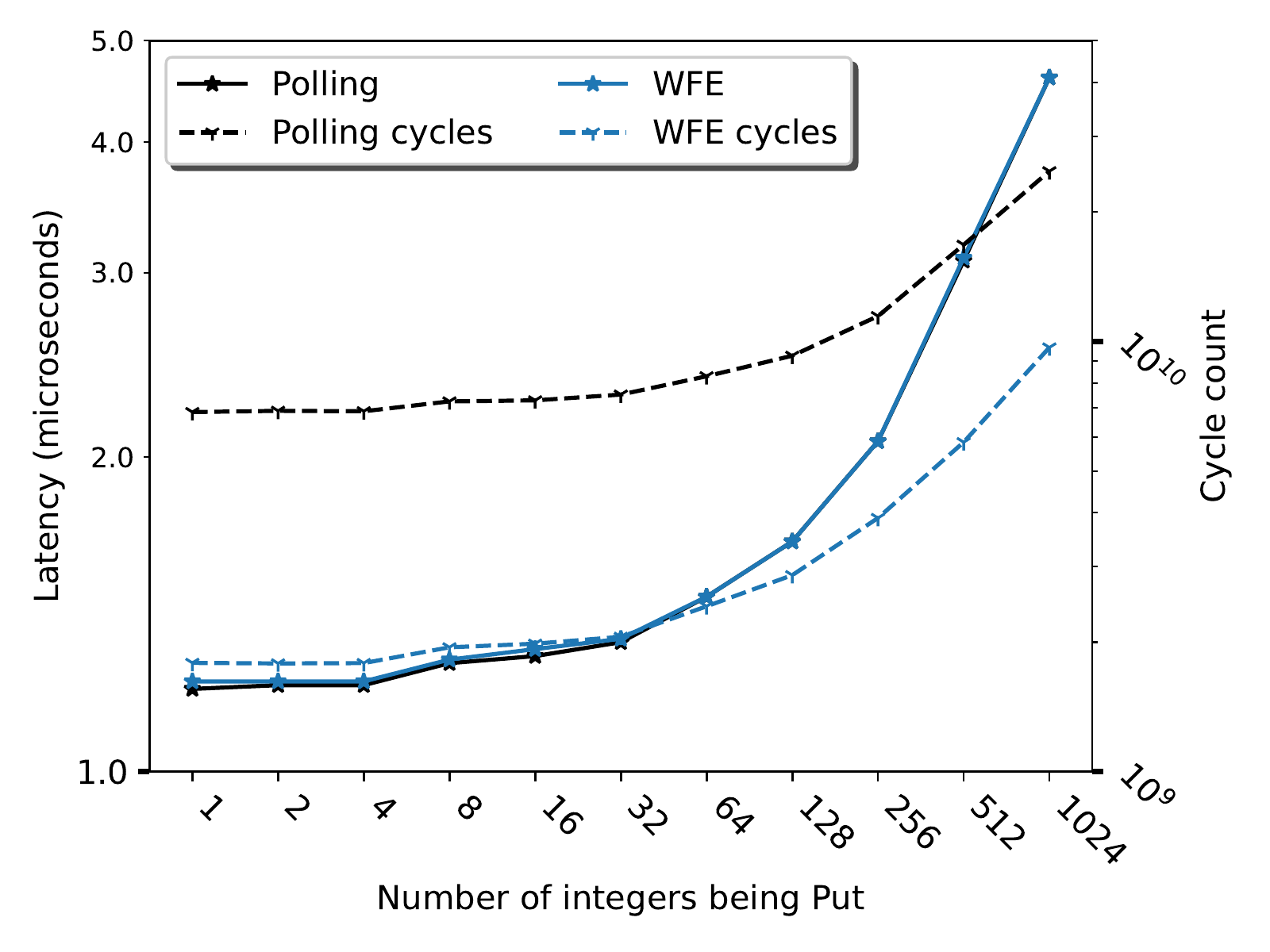}}
  \caption{\indput{}: Effects of using WFE on \cc{} active messages}
  \label{fig:wfe_ind}
\end{figure}

When running the \ssum{} active message (\figurename~\ref{fig:wfe_ssum}), there
is virtually no latency difference between busy-waiting (\textit{Polling}) and
using \textit{WFE}. There is a significant difference in cycle count between
\textit{Polling} and \textit{WFE}. When using the 512B message size, the
\textit{WFE} benchmark uses only 27\% of the cycles required by the
\textit{Polling} benchmark, a 3.6× reduction. For the 32KB message size, the
difference contracts to 1.84×.

\begin{figure}[tbp] 
  \centerline{\includegraphics[width=0.5\textwidth]{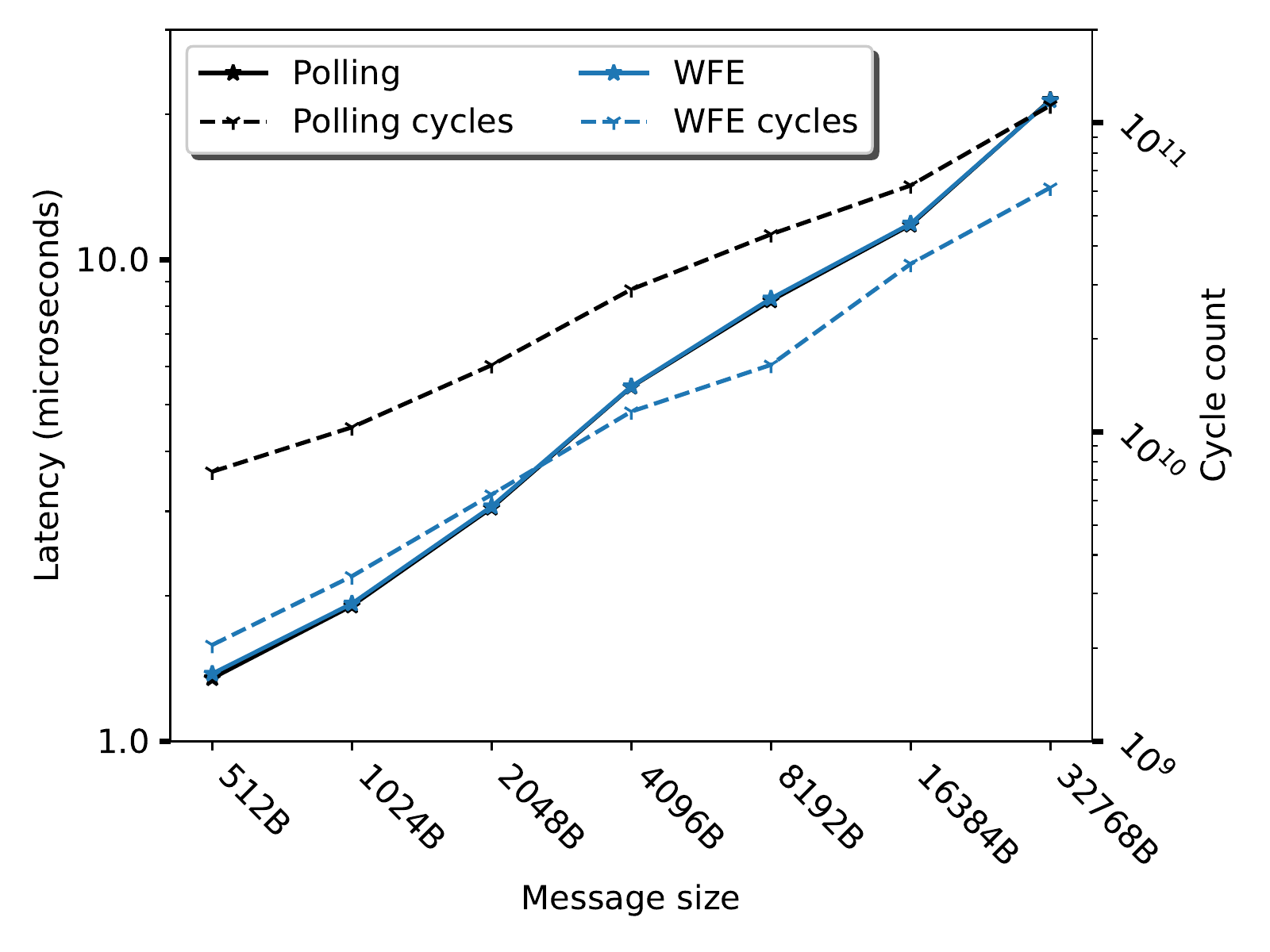}}
  \caption{\ssum{}: Effects of using WFE on \cc{} active messages}
  \label{fig:wfe_ssum}
\end{figure}
\section{Conclusion and Future Work}
In this work, we presented the \cc{} active message framework that implements a
new class of active message semantics where data and binary functions are
delivered over the RDMA network. We demonstrated how these semantics can be used
to implement advanced communication functions such as \ssum{} and \indput{}. The
framework uses remote dynamic linking and loading to resolve active messages to
external symbolic references for functions arriving over the network. \cc{}
injects binary functions and data to a remote system's last level cache over the
RDMA network. We demonstrated that this optimization reduces the \indput{}
active message latency by up to 31\% and increases its injection rate by up to 
92\%. In addition, this optimization reduces the tail latency by x2.4 for fully
loaded systems. To address efficiency concerns typically associated with busy
polling on message arrival, we used the \textit{WFE} instruction to reduce the
number of cycles spent on polling by up to 3.8x without sacrificing latency.

In future research we plan to extend \cc{} function injection logic to detect
reoccurring functions that have been injected and auto-switch to local function
execution while reducing the size of the active message. In addition, we plan to
integrate the \cc{} framework with the Charm++ UCX conduit~\cite{charmucx} for
further framework evaluation. On the security front, we will explore how efforts
in the confidential compute domain~\cite{confidentialcompute} can be used to
improve the security of function injection over an RDMA network.
\section{Acknowledgments}
The authors would like to thank the Los Alamos National Laboratory for their
continued support of this project. We thank Gilad Shainer from Nvidia for
providing us with the latest generation of Nvidia InfiniBand hardware. In
addition, we would like thank Jamshed Jalal, Eric Van Hensbergen, James Yang,
Ashwin Matta, and Anitha Kona from Arm for their support of this project.

\bibliographystyle{IEEEtran}
\bibliography{IEEEabrv,cluster21}

\end{document}